\documentclass[nologo,11pt,a4paper]{ETHpaper}
\usepackage{enumitem}

\tikzset{
    position/.style args={#1:#2 from #3}{
        at=(#3.#1), anchor=#1+180, shift=(#1:#2)
    },
    minimum size=10pt,
    every loop/.style={min distance=20mm,in=45,out=-45,looseness=10}
}

\newcommand{\model}[1]{#1}

\title{The likelihood-ratio test for multi-edge network models}
\titlealternative{The likelihood-ratio test for multi-edge network models}
\author{Giona Casiraghi}
\address{ETH Z\"urich, Chair of Systems Design\\ Weinbergstrasse 56/58,  Z\"urich, Switzerland \\ gcasiraghi@ethz.ch}
\www{\url{http://www.sg.ethz.ch}}
\reference{(Submitted for publication)}

\begin{document}

\maketitle

\begin{abstract}

The complexity underlying real-world systems implies that standard statistical hypothesis testing methods may not be adequate for these peculiar applications.
Specifically, we show that the likelihood-ratio test's null-distribution needs to be modified to accommodate the complexity found in multi-edge network data.
When working with independent observations, the p-values of likelihood-ratio tests are approximated using a $\chi^2$ distribution.
However, such an approximation should not be used when dealing with multi-edge network data.
This type of data is characterized by multiple correlations and competitions that make the standard approximation unsuitable.
We provide a solution to the problem by providing a better approximation of the likelihood-ratio test null-distribution through a Beta distribution.
Finally, we empirically show that even for a small multi-edge network, the standard $\chi^2$ approximation provides erroneous results, while the proposed Beta approximation yields the correct p-value estimation.  
    
    \emph{Keywords: \textbf{likelihood-ratio test}, \textbf{multi-edge network}, \textbf{complex system}, \textbf{hypothesis testing}, \textbf{model selection}}
\end{abstract}

\section{Overview}

Complex systems are notoriously challenging to analyze due to the large number of interdependencies, competitions, and correlations underlying their dynamics.
To deal with these issues, data-driven studies of complex systems are based -- either directly or indirectly -- on the careful formulation of models representing different hypotheses about the system.
The validation of these hypotheses is performed by comparing how well different models fit some observed data $\G$.
Principled \emph{model selection} is most probably the central problem of data analysis.

Model selection and statistical hypothesis testing have been intensively investigated in the general cases of, e.g., linear and generalized statistical regression models~\cite{Lehmann3,Burnham2004}.
However, less attention has been devoted to developing hypothesis testing methods specific to network models and network data, commonly used to study complex systems.
In this article, we investigate how one standard statistical test, the \emph{likelihood-ratio (LR) test}, needs to be modified when dealing with \emph{multi-edge network data}.

Model selection is addressed by operationalizing the principle of parsimony, one of the fundamental concepts in statistical modeling.
Statisticians usually view the principle of parsimony as a \emph{bias versus variance tradeoff}:
bias decreases, and variance increases as the complexity of a model increases.
The fit of any model can be improved by increasing the number of parameters.
However, a tradeoff with the increasing variance must be considered when selecting a model to validate a statistical hypothesis.
Parsimonious models should achieve the proper tradeoff between bias and variance.
\Citet{Box1994} suggested that the principle of parsimony should lead to a model with ``the smallest possible number of parameters for adequate representation of the data".
Data-driven selection of a parsimonious model is thus at the core of scientific research.

We can roughly summarise model selection methods into two groups that address the principle of parsimony differently.
Model selection based on statistical tests and model selection based on information-theoretic methods~\cite{Burnham2004}.
Prominent examples of information-theoretic methods are the AIC~\cite{Akaike1973,Akaike1974}, the BIC~\cite{Schwarz1978}, or description length minimisation~\cite{infomap,blockmodel}.
Studying how such methods fare when faced with network data complexity is beyond this article's scope.

The LR test is instead one of the most common examples of statistical tests used for hypothesis testing.
Statistical tests allow performing hypothesis testing in the following way.
They evaluate how far a \textbf{test statistic} $\lambda$ falls from an appropriately constructed null-model.
In the LR test, the test statistic $\lambda$ is the ratio between the likelihood $L_0$ of the model $X_0$ representing the null-hypothesis and the likelihood $L_a$ of a more complex model $X_a$ representing the alternative hypothesis to be tested.
The better the alternative hypothesis is compared to the null, the smaller the test statistic's absolute value $\lambda$.

How small need $\lambda$ be to reject the null hypothesis in favor of the alternative?
This depends on the null-distribution of $\lambda$.
In other words, it depends on the null-hypothesis and its corresponding null-model $X_0$.
Assuming that the null-hypothesis is correct, we could generate realizations $\tilde\G$ of model $X_0$ that represent all the possible forms the null-hypothesis could have taken in the data.
This provides a \emph{null-distribution} for the test statistic, i.e., a baseline distribution of the test statistic assuming that the null hypothesis was true.
If the alternative hypothesis does not fit the observed data well, we can expect the probability of observing $\lambda$ from the null-distribution to be relatively large.
The reason for this is that $X_a$ does not fit the data better than $X_0$.
In other words, there is no (statistical) evidence that we need the more complex model $X_a$ to explain the data, and the null-model $X_0$ is sufficient.
If the alternative hypothesis is considerably better than the null, the probability of observing $\lambda$ under the null-hypothesis will be small.
This would give statistical evidence to reject the null hypothesis in favor of the alternative.
The p-value of the LR test is precisely the probability of observing a value from the null distribution as small or smaller than $\lambda$.

Standard implementations of the LR test have been developed assuming that the data consist of many independent observations of the same process (i.i.d observations)~\cite{Lehmann3}.
Under these circumstances, Wilk's theorem provides a widely used approximation of the null-distribution of $\lambda$ to a $\chi^2$ distribution~\cite{Rao1973}.
Analyzing complex systems, we are often faced with \emph{multi-edge network data}.
These data consist of $m$ repeated -- and possibly time-stamped -- edges $(i,j)$ representing interactions between $n$ different agents $i,j$, the vertices of the network.
Examples of such datasets arise in multiple fields, e.g., in the form of human or animal interactions.
Usually, events in networks are explicitly dependent on each other.
Thus, the crucial assumption of i.i.d observations required by Wilk's theorem is violated.
The underlying assumption in network science is that the opposite of i.i.d. is true, i.e., that the presence or absence of edges between some vertex pairs affects edges between other vertex pairs.
This is critical in the presence of phenomena typically found in complex social systems, such as triadic closure~\cite{triadic}, structural balance~\cite{balance}, degree-degree correlations~\cite{degree-degree}, and other network effects.

So what is the implication of such interdependencies?
The dependence between different observations of a complex system means that some of the statistical tests' properties will not hold when analyzing network data.
In particular, we show that the null-distribution of the test statistic $\lambda$ of the LR test needs to be modified to accommodate such dependence.
When this is not done, the results obtained applying LR tests for hypothesis testing cannot be relied upon.

To illustrate this, we employ the gHypEG (generalized hypergeometric ensemble of random graphs)~\cite{Casiraghi2018,Casiraghi2017} to model multi-edge network data.
The gHypEG allows the encoding of different types of hypotheses in a model, from simple ones like block structures~\cite{Casiraghi2018a} to more complex ones, akin to statistical regression models~\cite{Casiraghi2017b,Brandenberger2019qauntifying}.
Such models can then be used to evaluate different hypotheses about the data~\cite{brandenberger2021onlineoffline}.

\section{Statistical Hypotheses and gHypEG}

\subsection{Multi-Edge Network Data and Hypothesis Formulation}

In this article, we deal with multi-edge network data.
Each observation $e_{i\to j}$ consists of an interaction from a system's agent $i$ to another agent $j$.
All observations $E=\{e_{i\to j}\}$ can be collected as directed edges in a multi-edge network $\G(V,E)$, where $V$ is the set of all interacting agents, the vertices of the network.
The matrix $\pmb A$ denotes the adjacency matrix of the network $\G$.
Each of its entries $A_{ij}$ reports the number of edges from vertex $i$ to vertex $j$, or in other words, the number of observed interactions between agent $i$ and agent $j$.

Within this framework, statistical hypotheses are formulated in terms of \emph{random graph models}.
Simple examples of such hypotheses are:
\begin{enumerate}[label=(\alph*)]
  \item Each agent has the same potential to interact as any other.
  \item Different agents have different potentials to interact.
  \item Agents are separated into 2 distinct groups, and agents of one group are more likely to interact with each other than with agents of the other group.
  \item Agents are separated into $n$ distinct groups, and agents of one group are more likely to interact with each other than with agents of the other groups.
\end{enumerate}
More complex social and dynamical hypotheses can obviously be formulated depending on the system studied.
Testing these hypotheses requires encoding them in statistical models and then comparing the fit of these models against each other or against some null-models~\cite{Lehmann3}.
We employ the generalized hypergeometric ensemble of random graph (gHypEG) as the statistical model encoding such hypotheses to deal with multi-edge network data.
While this is not the only option, we choose the gHypEG because of its versatility and suitability to model multi-edges.
In the next sections, we show i) how to formulate a likelihood-ratio test between two of such hypotheses and ii) how the null-distribution of the LR test needs to be modified to fit complex network data.

\subsection{The Generalised Hypergeometric Ensembles of Random Graphs (gHypEG)}

Before discussing the LR test details, we briefly introduce the generalized hypergeometric ensemble of random graphs.
A more formal presentation is provided in~\citep{Casiraghi2018}.

\begin{figure}[h!]
\centering
\includegraphics[width=.32\textwidth]{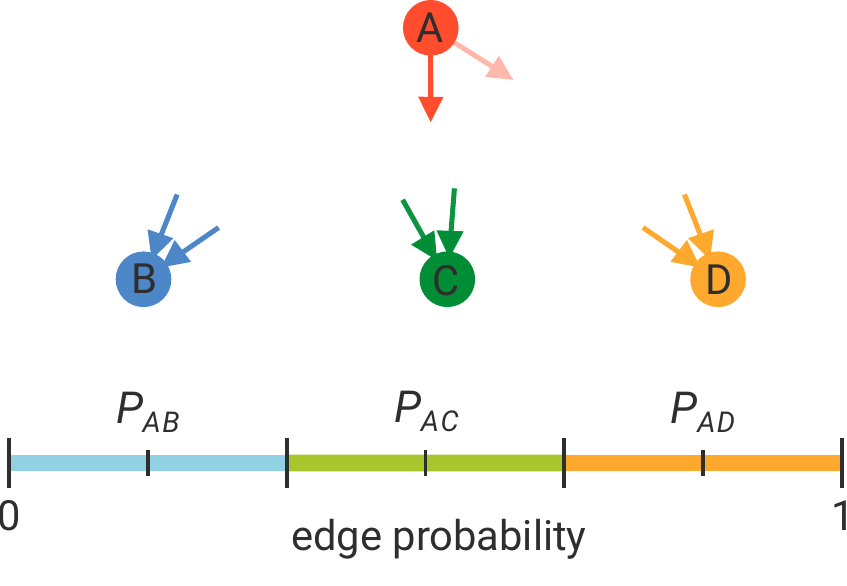}
\hfill
\includegraphics[width=.32\textwidth]{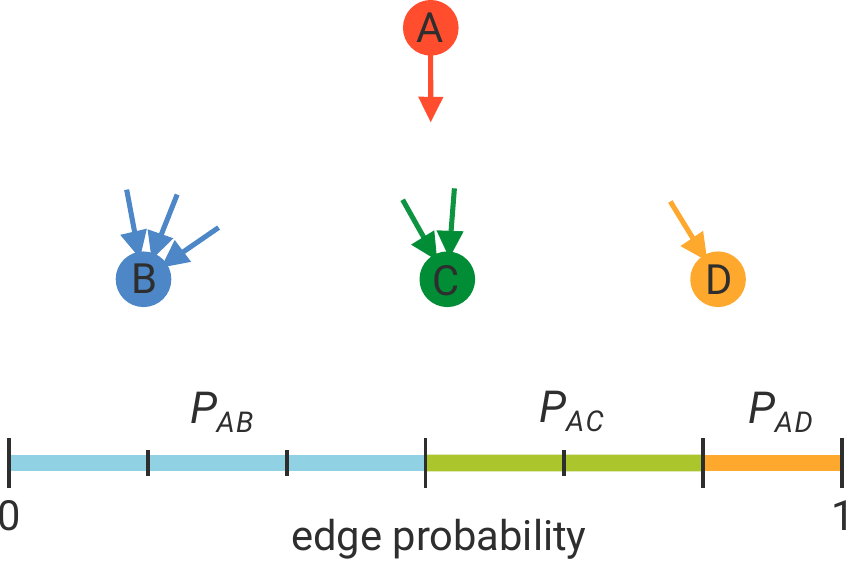}
\hfill
\includegraphics[width=.32\textwidth]{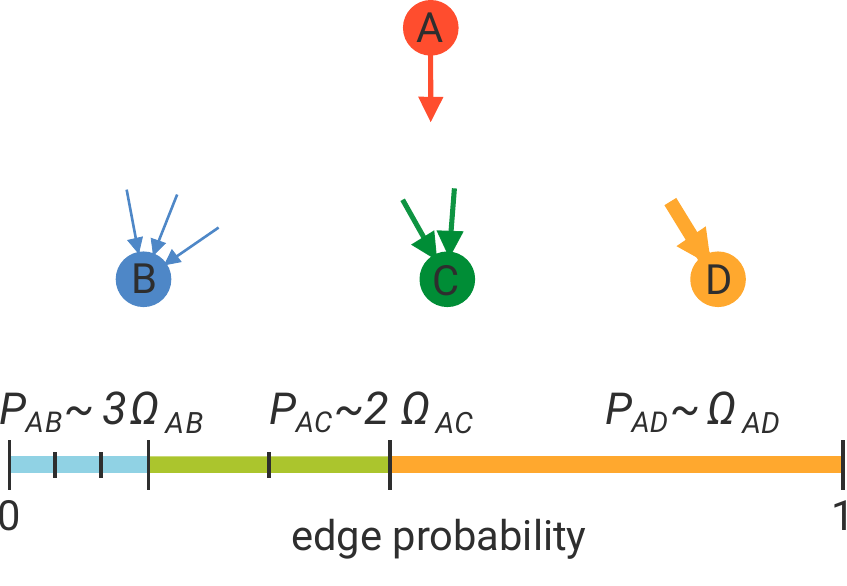}
\caption{Probabilities of connecting different agents according to gHypEG, from the perspective of an agent A.
Graphical illustration of \textbf{(left)} uniform edge probabilities, \textbf{(center)} the probability of connecting two vertices is a function of degrees, and \textbf{(right)} the probability of connecting two vertices is function of degrees and propensities $\Omega$ (represented as the edge-weight).}
\label{fig:props}
\end{figure}

The general idea underlying the gHypEG is to sample edges at random from a predefined set of possible edges.
Hypotheses about the system from which the data is observed can be encoded by either a) changing the number of possible edges in such a set and b) changing the odds of sampling an edge between two vertices instead of others, i.e., by specifying different edge sampling weights, or biases.
\Cref{fig:props} provides a graphical illustration of such a process from the perspective of an agent A.

On the left-hand side of \cref{fig:props}, the number of possible ways A can interact with the other agents is the same: there are two \emph{edge-stubs} for each vertex.
Moreover, the odds of sampling one edge-stub instead of another is $1$.
I.e., each edge has the same sampling weight, which is denoted as $\Omega_{ij}$ in the following, where $i,j$ are vertices in $V$.
According to the model just described, the probability of observing a multi-edge network $\G$ with $m$ edges depends only on the number $\Xi$ of possible edges between each pair of vertices.
This scenario gives rise to a uniform random graph model similar in spirit to the $G(n,p)$ model of Erd\"os and R\`enyi~\cite{Erdos1959}.
The process of sampling $m$ edges from a collection of $n^2\Xi$ possible edges, i.e, $\Xi$ possible edges for each pair of vertices in a directed network with selfloops, is described by the hypergeometric distribution~\cite{Casiraghi2018}:
\begin{equation}\label{eq:hyper_flat}
    \Pr(\G\lvert\Xi)=\dbinom{n^2\Xi}{m}^{-1}\prod_{i,j\in V}{\dbinom{\Xi}{A_{ij}}},
\end{equation}
By setting $\Xi=(m/n)^2$, we ensure that the \emph{average degree} of the observed network $\G$ is preserved by the model.
This first scenario corresponds to the hypothesis (a) listed above: each agent has the same potential to interact.
Furthermore, in a directed network with self-loops, $\Xi=m^2/n^2$ corresponds to the maximum likelihood estimation of the only model parameter $\Xi$.
We refer to the resulting hypergeometric network model as \emph{regular model}.

The central illustration of \cref{fig:props} highlights a different case.
The odds between the different interactions are still identical.
I.e., there is no preference for A to interact with any of the other agents.
However, the actual possibilities of interactions vary between the different agents:
each agent has a different number of edge-stubs for A to connect to.
This scenario encodes a different potential of interaction for the different agents, usually reflected in a heterogeneous degree distribution found in the network $\G$.
This model encodes hypothesis (b) above.
In practice, this hypothesis requires setting different values $\Xi_{ij}$ for the number of possible edges between each pair vertices $i,j$.
The probability of observing a network $\G$ according to this model changes as follows:
\begin{equation}\label{eq:hyper}
    \Pr(\G\lvert\pmb\Xi)=\dbinom{\sum_{ij}\Xi_{ij}}{m}^{-1}\prod_{i,j\in V}{\dbinom{\Xi_{ij}}{A_{ij}}},
\end{equation}
where the matrix $\pmb\Xi$ contains all different entries $\Xi_{ij}$.
The value of $\Xi_{ij}$ can be freely chosen to encode different properties of the system studied (see, e.g.,~\cite{brandenberger2021onlineoffline}).
For example, if we were studying a citation network consisting of citations between scientists, we could set $\Xi_{ij}$ to $p_i\cdot p_j$, where $p_x$ is the number of articles published by scientist $x$.
$p_i\cdot p_j$ would then encode all the possible ways scientist $i$ \emph{could have} cited scientist $j$, through all their respective publications.
In most cases, though, $\Xi_{ij}$ is taken to be $k_i^\text{out}\cdot k_j^\text{in}$, where $k_x^\text{in}$ is the observed in-degree of agent $x$, and $k_x^\text{out}$ its observed out degree.
This hypergeometric network model corresponds to a soft \emph{configuration model}~\cite{fosdick2018}, and defines a network model that preserves the observed degree sequences in expectation~\cite{Casiraghi2018}.

The two models described so far are both characterized by the absence of sampling biases, i.e., interaction preferences between specific vertex pairs that go beyond what is prescribed by the number of edge-stubs and degrees.
GHypEG further expands this formulation modifying the hypergeometric configuration model with additional information available about the system.
Specifically, the probability of connecting two vertices depends not only on the observed degrees (i.e., number of stubs) but also on an independent \emph{propensity} of two vertices to be connected.
Such propensities introduce non-degree related effects into the model.
This result is achieved by changing the \emph{odds} of connecting a pair of vertices instead of another.
The right side of \cref{fig:props} illustrates this case, where $A$ is most likely to connect with vertex $D$, even though $D$ has only one available stub.

We collect these edge propensities in a matrix $\bm\Omega$.
The ratio between any two elements $\Omega_{ij}$ and $\Omega_{kl}$ of the propensity matrix gives the odds-ratio of observing an edge between vertices $i$ and $j$ instead of $k$ and $l$, independently of the degrees of the vertices.
The probability of a graph $\G$ depends on the stubs' configuration specified by $\bm\Xi$, and on the odds defined by $\bm\Omega$.
Such a probability distribution is described by the multivariate Wallenius' non-central hypergeometric distribution~\citep{wallenius1963, Chesson1978}:
\begin{equation}\label{eq:hyperI}
\Pr(\G|\bm\Xi,\bm\Omega)=\left[\prod_{i,j}{\dbinom{\Xi_{ij}}{A_{ij}}}\right]
         \int_{0}^{1}{\prod_{i,j}{\left(1-z^{\frac{\Omega_{ij}}{S_{{\Omega}} }}\right)^{A_{ij}}}dz}
\end{equation}
with $S_{{\Omega}}= \sum_{i,j} \Omega_{ij}(\Xi_{ij}-A_{ij})$.

\subsection{Encoding Hypotheses}

By constraining the number of free parameters in $\pmb\Omega$, we can specify hypotheses about the data changing the sampling odds for different vertex pairs.
For example, we can cluster vertices into multiple groups and verify whether the odds of observing interactions \emph{within a group} and \emph{between a group} are different~\cite{Casiraghi2018a}.
The resulting model is similar to a degree-corrected stochastic block model~\cite{Karrer2011}.
Alternatively, we can specify $\pmb\Omega$ to encode endogenous network properties.
E.g., $\pmb\Omega$ can be utilized to encode \emph{triadic closure}~\cite{Brandenberger2019qauntifying}, to verify whether pairs whose interactions will close triads in the network are more likely than others.
Finally, different effects contributing to the odds of observing some interactions instead of others can be composed together to formulate more complex hypotheses~\cite{Casiraghi2017b}.

The advantage of the approach just described is the ability to encode a wide range of statistical hypotheses within the same modelling framework.
This has the practical benefit of allowing the comparison of very different models, as they can all be formulated by means of the same probability distribution of \cref{eq:hyperI}.
Different hypotheses are thus encoded by appropriately choosing the free parameters in $\pmb\Xi$ and $\pmb\Omega$.

For clarity, we will focus on simple hypotheses such as those described in the previous section.
However, the results shown do hold for any combinations of $\pmb\Xi$ and $\pmb\Omega$.
In the particular case of encoding group structures, $\pmb\Omega$ takes the following form:
\begin{equation}
  \Omega_{ij} := \omega_{g_i,g_j},
\end{equation}
where $g_i$ is the group of agent $i$, $g_j$ is the group of agent $j$, and $\omega_{g_i,g_j}$ is the propensity of sampling an edge between group $g_i$ and group $g_j$.
In the presence of 2 different groups of vertices $(A,B)$, there are 3 possible values that $\Omega_{ij}$ can take:
$\omega_{AA},\,\omega_{BB},\,\omega_{AB}$ (assuming that $\omega_{AB} = \omega_{BA}$).
The ratio $\omega_{AA}/\omega_{AB}$ gives the odds between sampling an edge within group A and an edge between group A and B given a value of $\Xi$.

\section{The Likelihood-ratio (LR) test}\label{sec:lrtest}

We now illustrate how the LR test is used to test a null-hypothesis against an alternative hypothesis about the observed system.
The data are used to define a graph $\G$ with adjacency matrix $\bm{A}$.
Let $H_r$ be some statistical hypothesis.
Here, we always assume that each hypothesis is defined by a gHypEG model $X_r$ that can be encoded by a propensity matrix $\bm\Omega_r$ and a combinatorial matrix $\bm\Xi_r$.
Each model is characterized by several free parameters that we want to fit to the data $\G$, such that the probability of observing $\G$ is maximized.
This requirement corresponds to performing a maximum likelihood estimation (MLE) of the free parameters.

\paragraph{Likelihood-ratio statistic}
Assume now we have two hypotheses we want to test against each other.
Let $H_0$ denote the null-hypothesis and let $H_a$ denote the alternative.
The corresponding models are defined in terms of $\bm\Omega_0$, $\bm\Xi_0$ and $\bm\Omega_a$, $\bm\Xi_a$.
To test the alternative hypothesis against the null, we use the likelihood-ratio statistic $\lambda(0,a)$, defined as follows:
\begin{definition}[Likelihood-ratio statistic]
Let $\G$ be a graph, $\model X_0$ be the model corresponding to the null-hypothesis, and $\model X_a$ the model corresponding to the alternative hypothesis.
The likelihood ratio statistic $\lambda(0,a)$ is given by
    \begin{equation}
  \lambda(0,a) := \frac{L(\bm\Xi_0,\bm\Omega_0\vert\bm A )}{\sup(L(\bm\Xi_0,\bm\Omega_0\vert\bm A ),L(\bm\Xi_a,\bm\Omega_a\vert\bm A ))},
\end{equation}
where $L(\bm\Xi_r,\bm\Omega_r\vert\bm A ) = \Pr(\G\vert\bm\Xi_r,\bm\Omega_r)$ denotes the likelihood of model $X_r$ given the network $\G$. 
\end{definition}

Through the likelihood-ratio statistic, we can perform two types of tests.
First, we can perform a standard model selection test to compare a simpler model against a more complex model.
This test corresponds to verify whether there is enough evidence in the data that justifies the more complex model or whether the simpler model fits the data well enough.
In this scenario, the simple model corresponds to the null-hypothesis, while the more complex model to the alternative.

Second, the likelihood-ratio test can be used to perform a goodness-of-fit test.
This test allows verifying the quality of the fit of a model $X_r$.
By defining the alternative hypothesis with a model $X_\text{full}$ that perfectly reproduces the observed data (in expectation), we can test whether the fit of the model $X_r$ is as good as such an overfitting model~\cite{Lehmann3}.
In the framework of gHypEGs, the alternative hypothesis is obtained by specifying the parameter matrix $\pmb\Omega_\text{full}$ such that the expectation of $X_\text{full}$ corresponds to the observation $\G$.
This model is the maximally complex model that can be specified with a gHypEG and has as many free parameters as entries in the adjacency matrix~\cite{Casiraghi2018}.

Let's now assume that the two models corresponding to the alternative and null hypotheses are \emph{nested}.
This means that $\bm\Xi_0$ can be written as a special case of $\bm\Xi_a$, and $\bm\Omega_0$ as a special case of $\bm\Omega_a$.
Thus, the null-model (with fewer parameters) can be formulated by constraining some of the alternative model parameters.
Thanks to Wilks' theorem~\cite{Rao1973}, if the two models are nested, the number of observations $m$ is large, and the observations are independent,
 the distribution of $\lambda$ under the null-hypothesis can be written in terms of
\begin{equation}\label{eq:dlambda}
    D(0,a):=-2\log(\lambda(0,a)),    
\end{equation}
and can be approximated by the $\chi^2$ distribution with as many degrees of freedom as the difference of degrees of freedoms between the two models.
Letting $\nu$ be the difference of degrees of freedom between the null and the alternative modes, the $p$-value of the likelihood-ratio test between the two hypotheses is computed as follows:
\begin{equation}
\text{p-value} := \Pr(\chi^2(\nu)\geq D(0,a)).
\label{eq:pvalue}
\end{equation}
We reject the null hypothesis in favour of the alternative if the $p$-value is smaller than some threshold $\alpha$.

\paragraph{Distribution of $\lambda$ under the null-hypothesis}
The question that remains to be answered is whether the conditions provided by multi-edge network data allow Wilks' theorem's application.
Unfortunately, in most real-world scenarios, the answer to this question is negative.
This is a known issue in statistics, where it arises in the context of multinomial goodness-of-fit tests~\cite{Smith1981,Chapman1976,Koehler1980,Larntz1978}.
Because Wallenius' multivariate non-central hypergeometric distribution converges to the multinomial distribution (a formal proof can be found in \cite{Zingg2019}), we use the results obtained for multinomial tests to find a better approximation for the null distribution of $D(0,a)$ than the $\chi^2$ approximation of Wilks' theorem.
Specifically, following the work of \citet{Smith1981}, we propose approximating the distribution of $D(0,a)$ with a Beta distribution.
\begin{thm}[Convergence in distribution of likelihood-ratio statistics]\label{thm:beta}
    The distribution under the null-hypothesis of $D(0,a)$, defined as in \cref{eq:dlambda}, for $m$ large tends towards a Beta$(\alpha,\beta)$ distribution with parameters

    \begin{equation}
        \alpha=\frac{\mu\left[D(0,a)\right]}{M\cdot\sigma^2\left[D(0,a)\right]} \cdot 
        \left( 
        		\mu\left[D(0,a)\right] \cdot \left( M - \mu\left[D(0,a)\right] \right) - 
        		\sigma^2\left[D(0,a)\right]
        \right),
    \end{equation}
    and
    \begin{equation}
    \beta= \left( M - \mu\left[D(0,a)\right] \right) \cdot \frac{\alpha}{\mu\left[D(0,a)\right]},
    \end{equation}
    where $M$ denotes the upper limit of the image of $D(0,a)$, $\mu\left[D(0,a)\right]$ its expectation and $\sigma^2\left[D(0,a)\right]$ its variance.
\end{thm}
In some special cases there exist analytical solutions for $\mu\left[D(0,a)\right]$ and $\sigma^2\left[D(0,a)\right]$~\cite{Smith1981}.
However, in most situations, we resort to a numerical estimation of them.
While a general solution would be optimal, thanks to the ability to generate samples provided by gHypEG models, the parameters' numerical estimation can be nevertheless performed with ease.

\begin{figure}
\centering
\resizebox{\textwidth}{!}{\input{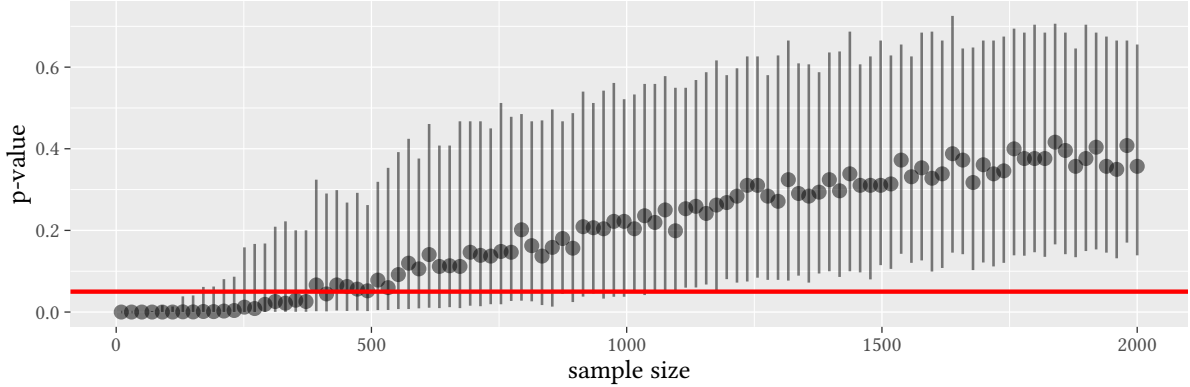}
}
\caption[P-values of a Kolmogorov-Smirnov test for the likelihood-ratio distribution.]{P-values of a Kolmogorov-Smirnov test for the likelihood-ratio null-distribution.
         Each point is the p-value (y-axis) of the median statistic resulting from 500 tests of the empirical distribution against the asymptotic Beta distribution, computed for an increasing sample size (x-axis).
         The length of the lines denotes 1.5 inter-quantile ranges.
         The horizontal red line shows the $0.05$ significance threshold for the Kolmogorov-Smirnov test.
         It is evident a clear trend where the p-value increases with increasing sample size.
         That highlights an increase of the goodness-of-fit of the limiting Beta distribution for the empirical distribution.
}\label{fig:ks_beta}    
\end{figure}

The result provided by \cref{thm:beta} greatly helps when performing likelihood-ratio tests involving multi-edge network data.
In fact, to estimate the full null distribution of the likelihood-ratio statistic numerically, we would need a considerable number of realisations.
In the case of large networks, this is infeasible.
Exploiting \cref{thm:beta} instead, we only need to estimate the first two moments of the distribution under the null hypothesis, which can be done reliably with a smaller number of realisation~\cite{Shore1995}.
For example, in \cref{fig:ks_beta}, we show the results of applying Kolmogorov-Smirnov's test to compare the LR's distribution statistics against the Beta distribution fitted to increasing sample sizes.
The example is constructed from a 40 vertices random graph with 500 undirected edges.
The edges are generated according to the hypergeometric configuration model, and the likelihood-ratio test is performed comparing a regular model against the generating configuration model.
To build the empirical distribution of the likelihood-ratio statistic, we take $S=500\,000$ samples under the null hypothesis, and we compute the parameters of the asymptotic Beta distribution from an increasing number of independent samples.
The results show that with a limited sample size $s\sim 1000$, most of the observations give a p-value for the Kolmogorov-Smirnov test larger than $0.05$.
I.e., with a limited sample size $s$, the empirical null-distribution obtained from $S$ is not significantly different from the Beta distribution whose parameters have been estimated from the $s$ realizations.
That points to the fact that the asymptotic results of \cref{thm:beta} are acceptable even for a finite number of observations and small sample sizes.
The \texttt{R} package \texttt{ghypernet}\footnote{\url{https://ghyper.net}}~\cite{Casiraghi2019} provides an implementation of the likelihood-ratio test for gHypEG models.
The package is Open Source and can be obtained from the CRAN \texttt{R} package repository.

\section{Case studies}
We provide two short case studies about the application of the likelihood-ratio test.
First, we generate a random undirected graph with $n=100$ vertices and $m=400$ directed edges uniformly distributed between each vertex pairs.
Utilizing the likelihood-ratio test, we can test the null-hypothesis (a) that each vertex has the same potential of interactions against the alternative hypothesis (b) that different agents have different interaction potentials.
As explained in the previous sections, (a) is encoded by a \emph{regular model} with one parameter, and (b) by a \emph{configuration model}.
This test corresponds to testing that the degree distribution deviates from that of the regular model.
We expect that the test returns high p-values because we choose the null-hypothesis to match the random graph's generating process.
The results, obtained from 1000 repetitions of the experiment, confirm this hypothesis with a p-value of the median $\lambda$ of $0.44$.
Similarly, we perform the same experiment generating a random undirected graph from the standard configuration model with a heterogeneous degree distribution.
To ease the comparison with the example above, we define it by a degree sequence sampled from a geometric distribution with mean chosen such that the expected number of edges in the graph is $m=400$.
This effectively corresponds to generating data according to the hypothesis (b).
In this case, we expect small p-values from the same test done before because the generating model of the data corresponds to the alternative hypothesis.
Repeating the experiment 1000 times, for the largest recorded $\lambda$ we obtain a p-value $<1e-20$.

The second experiment we perform is with an empirical graph.
We use Zachary's Karate Club (ZKC)~\cite{Zachary} as a test case.
ZKC consists of 34 vertices and 231 undirected multi-edges.
As with most empirical graphs, its degree sequence is skewed (empirical skewness is $1.456$).
Hence, we expect that the test we performed before -- comparing the null-hypothesis of a regular model against the hypergeometric configuration model -- should reject the null-hypothesis.
Performing such a likelihood-ratio test gives a p-value $<1e-20$, which confirms our expectations.

\begin{figure}
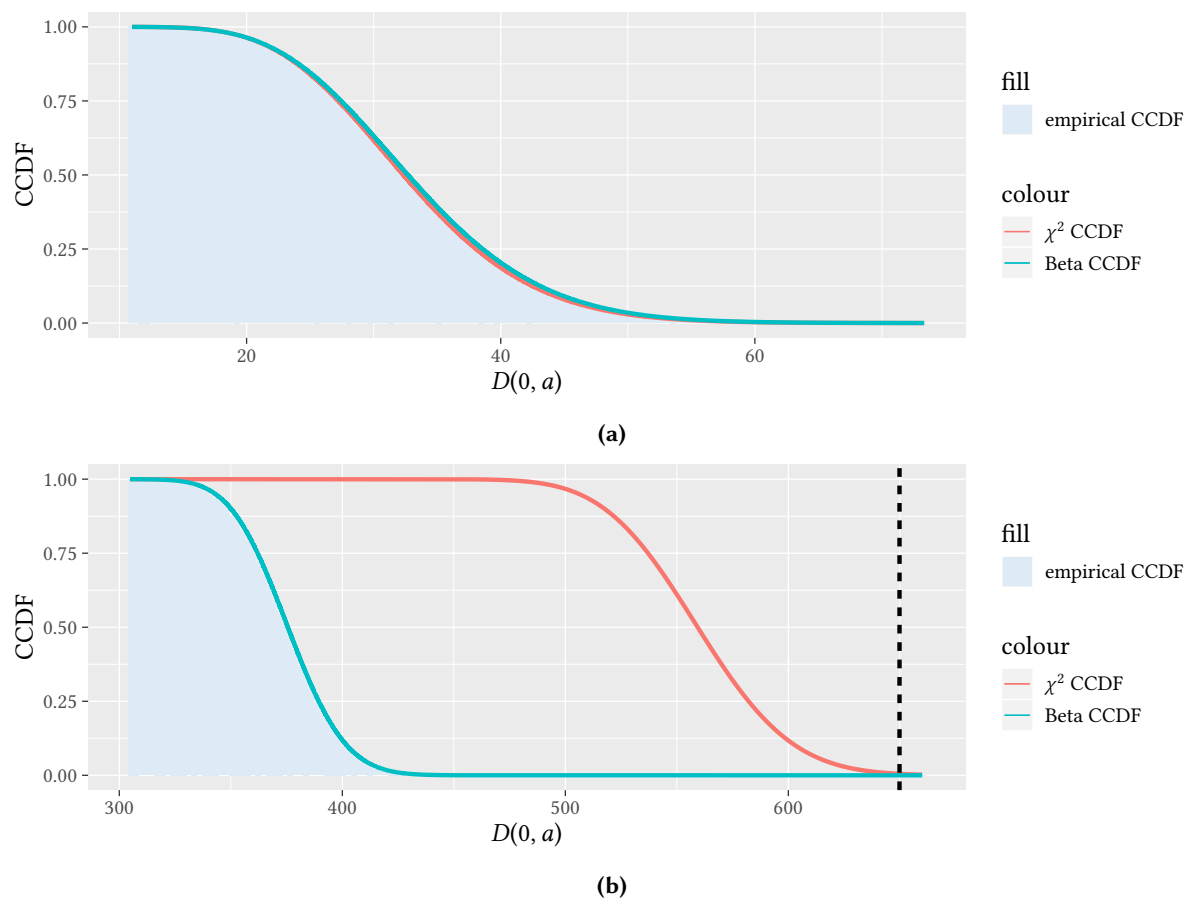

\begin{subfigure}{\textwidth}
\centering
\resizebox{\textwidth}{!}{\input{figures/rmvscm_karate}
}
\caption{}\label{fig:rmvscm}    
\end{subfigure}\\
\begin{subfigure}{\textwidth}
\centering
\resizebox{\textwidth}{!}{\input{figures/cmvsfull_karate}
}
\caption{}\label{fig:cmvsfm}
\end{subfigure}
\caption[Empirical distribution of $D(0,2)$ for two likelihood-ratio tests perform on ZKC.]
{Empirical distribution of $D(0,2)$ for two likelihood-ratio tests perform on ZKC.
         In the top figure, we perform a likelihood-ratio test for the null-hypothesis that ZKC comes from a regular model against the alternative hypothesis that ZKC comes from a configuration model.
         The results show that there is strong evidence to reject the null hypothesis.
         In the bottom figure, we perform a goodness-of-fit test for the hypergeometric configuration model on ZKC.
         Also, in this case, the results show a bad fit of the null-model.
         While in the top-figure, the $\chi^2$ and the Beta approximations of the likelihood-ratio statistic's empirical distribution give a relatively good fit, in the bottom case, it is clear that the $\chi^2$ does not approximate well the empirical distribution.
         The shaded area denotes the empirical distribution of $D(0,2)$ under the null-hypothesis, the orange line its $\chi^2$ approximation, and the green line its Beta approximation.
         The vertical line denotes the value of the likelihood-ratio statistic for ZKC.
         In the top figure, such a line is out of the boundaries on the right side and hence not plotted.}\label{fig:emplrtest}
\end{figure}

We can further exploit this example to compare the empirical distribution of $D(0, a)$ under the null hypothesis with the $\chi^2$ distribution and the Beta distribution.
The result is shown in \cref{fig:rmvscm}.
Note that the value of $D(0,a)$ for ZKC is $300.338$, which is out of scale on the right side of the x-axis of \cref{fig:rmvscm}.
In this case, it appears that the Beta distribution and the $\chi^2$ distribution provide similar fits.
However, when we perform a two-sided Kolmogorov-Smirnov test, we get a p-value of $1.45e-05$ for the $\chi^2$ distribution, which means that we can reject the null hypothesis that the empirical distribution follows a $\chi^2$.
Performing the same test for the Beta distribution, we get a p-value of $0.4211$.
That means that we cannot reject the null-hypothesis that the empirical distribution follows a Beta.
While this is not enough to claim that the distribution of $D(0, a)$ \textbf{is} a Beta, it gives confidence on using the asymptotic results of \cref{thm:beta}.

We perform a second experiment to highlight how much, in extreme cases, the $\chi^2$ distribution can deviate from the empirical distribution of $D(0, a)$ under the null-hypothesis.
For the ZKC, we now perform a goodness-of-fit test of the hypergeometric configuration model.
The alternative hypothesis is thus encoded by the maximally complex model fitted by gHypEGs and results in a model that fixes the expected graph as the observed one, as explained above.
The fit of its parameters is performed according to what described in \cite{Casiraghi2018}.
The test can be interpreted as how well the null-model fits the data, which is entirely encoded in the full model.
This test results in a p-value of $1.69e-30$.
That means that the configuration model is not a good model for the ZKC.
This result is hardly surprising, given the well-known community structure present in the empirical graph.
In \cref{fig:cmvsfm}, we show the empirical distribution of $D(0,a)$.
There, we notice that while the Beta distribution provides a visually good fit, the $\chi^2$ distribution is heavily shifted to the right.
The two-sample Kolmogorov-Smirnov test confirms this result, providing a p-value of $0.169$ for the Beta distribution, and $<2.2e-16$ for the $\chi^2$ distribution.

In this last example, it is essential to note that using the $\chi^2$ distribution would provide a misleading result.
Comparing the value of $D(0,a)$ for the empirical graph, we see that it is on the right tail of the $\chi^2$ distribution.
Computing a p-value from this distribution would result in a p-value of $\approx 0.005$.
That means that in this case, we would only weakly reject the null-hypothesis, giving the wrong impression that the ZKC could come from an extreme realization of a simple configuration model.
However, this is ruled out by looking at the likelihood-ratio statistics' empirical distribution or simply comparing an empirical graph with a realization from the hypergeometric configuration model~\cite{Casiraghi2016}.

\section{Discussion}

The study of complex systems is intertwined with network science and advanced multivariate statistics.
Hypothesis testing and model selection methods, in particular, need to account for the complexity underlying observations from such systems.
Because interactions between system agents tend not to be independent, many standard statistical methods should be employed with care when dealing with network data.

This article has investigated how the likelihood-ratio test needs to be adapted to deal with network models.
Likelihood-ratio tests provide a practical methodology for selecting different network models and testing statistical hypotheses.
However, the characteristics of multi-edge networks require us to adapt the test null-distribution to account for the underlying complexity of network data.
When this is not done, we incur the risk of over- (or under-) estimating the p-values of the statistical test, generating contradictory results, as shown in the case study above.
With \cref{thm:beta}, we provide the means to correctly estimate the p-values for likelihood-ratio tests by means of a Beta distribution.
Finally, we provide an implementation of the methods described through the Open Source \texttt{R} package \texttt{ghypernet}.
Even though our analysis is focused on the likelihood-ratio test, similar issues may arise with other statistical tools applied to complex networks.

The main limitation of the results presented in this article is the need to numerically estimate the first two empirical moments of the statistic's null distribution.
Although this can be performed easily using our implementation, we will investigate analytical asymptotic estimates for the parameters needed in future research.

\section*{Acknowledgements}

The author thanks Frank Schweitzer for his support and Georges Andres for his detailed comments.

\small \setlength{\bibsep}{1pt}

\end{document}